\documentstyle[seceq,preprint,epsf]{jpsj}

\renewcommand\figureheight[1]{\vspace{24pt}\mbox{\rule{0cm}{#1}}}

\def\runtitle{Rough Surface Effect on Meissner Diamagnetism
in Normal-layer of N-S Proximity-Contact System}
\def\runauthor{Kotaro {\sc Yamada}, Seiji {\sc Higashitani}
and Katsuhiko {\sc Nagai}}

\title
{
Rough Surface Effect on Meissner Diamagnetism
in Normal-layer of N-S Proximity-Contact System
}

\author
{
Kotaro {\sc Yamada}\footnote{Present address: Institute for Research
in Humanities, Kyoto University, Kyoto 606-8501.}, 
Seiji {\sc Higashitani}$^{1}$
and Katsuhiko {\sc Nagai}$^{1}$
}

\inst
{
Department of Engineering, Hiroshima University, 
Higashi-hiroshima 739-8527\\
$^1$Faculty of Integrated Arts and Sciences, Hiroshima University,
Higashi-hiroshima 739-8521
}

\recdate
{
August 31, 2001}

\abst
{
Rough surface effect on the Meissner diamagnetic current in 
the normal layer of
proximity contact N-S bi-layer is investigated in the clean limit.
The diamagnetic current and the screening length  
are calculated by use of quasi-classical Green's function.
We show that the surface roughness has a sizable effect, 
even when a normal layer width 
is large compared with the coherence length $\xi =v_{\rm F}/\pi T_{\rm c}$.
The effect is as large as that of the impurity scattering and also
as that of the finite reflection at the N-S interface. 
}

\kword
{
Meissner effect, proximity effect, rough surface,  screening length,
quasi-classical Green's function
}

\begin{document}
\sloppy
\maketitle

\section{Introduction}
A normal metal in contact with a superconductor is 
known to bear superconducting properties even if the normal metal 
has no pairing interaction.
The superconducting order in the normal layer induced by the proximity effect
produces
a finite diamagnetic current to expel the magnetic field. The
quantity of interest is the screening fraction, i.e., how great
part of the normal layer contributes to the Meissner effect.
Theories of diamagnetic response of N-S proximity contact
system in both the dirty and the clean limit have been already 
reported.\cite{orsay,deutcher,zaikin,narikiyo,higashitani,belziga,belzigb,belzigc}
While  early  experimental results\cite{oda1,oda2} were found
to be in agreement with the dirty limit theory,
some data of more recent experiments\cite{belzigc,oda3,mota,visani,oda4}
failed to be described satisfactorily by either the clean or
the dirty limit theory.

Belzig {\it et al}.\cite{belzigb} proposed a quasi-classical
theory of diamagnetic response  of a normal layer with impurities 
and pointed out that, even in 
such a clean system that has a longer mean free path than the layer
width, temperature dependence of the screening fraction can deviate 
considerably from that of the clean limit theory.
M\"uller-Allinger {\it et al}.\cite{belzigc}
tried to fit the quasi-classical theory\cite{belzigb} with 
their experimental data
using the mean free path $l_{\rm N}$ as a fitting parameter.
Although they found fairy good agreement in the temperature dependence of the
screening fraction with the experimental data of the samples
with various impurity concentration, there remain 
some samples which exhibit considerably different behavior that
cannot be fitted by the theory of ref.\ \citen{belzigb}.
Moreover, best fits were obtained by using the mean free path 
a few times smaller than the measured value.\cite{belzigc}

In the theory by Belzig {\it et al}.\cite{belzigb}, it is assumed that 
the N-S interface is transparent, i.e., electrons pass through 
the interface without reflection. It is also assumed that the
surface scatters electrons specularily. The effect of the
finite reflection at the interface was examined by Higashitani
and Nagai\cite{higashitani} for  clean systems and by
Hara {\it et al}.\cite{hara} for systems with impurities. 
They showed that the finite electron reflection
at the N-S interface 
has strong influence on the screening fraction.

The purpose of this paper is to study the effect of surface roughness
on the screening fraction using the quasi-classical Green's
function method.\cite{AAHN,NHYN,NYN}
The proximity induced order parameter extends from the N-S interface
into the normal layer with the length scale of the 
N side coherence length ${\xi}_{\rm N}(T)$. At low temperatures, it will
reach the end surface of the normal layer and feel the surface
roughness. It is expected, therefore, the surface roughness effect
is important at low temperatures.
To focus on the surface roughness effect, we study a clean system
and neglect the effect of finite reflection at the interface.
We show that the surface roughness has a quantitatively
large effect on the screening fraction.

This paper is organized as follows. In the next section, we derive
a linear response of the quasi-classical Green's function to the
external vector potential. In \S 3, we briefly review the
quasi-classical theory of rough surface effect by 
Nagato {\it et al}.\cite{NHYN,NYN}. In \S 4, explicit form of the
quasi-classical Green's function in the normal layer of our model
system is given. Diamagnetic current and the screening fraction
are discussed in \S 5 and 6. The last section is devoted
to summary and discussion. Throughout this paper, we use the
unit $\hbar=k_{\rm B}=1$.

\section{Quasi-classical Theory of Diamagnetic Current}
In this paper, we consider a clean N-S 
double layer system as shown in Fig.~\ref{fig:NS}.
The end wall at $z=0$ of the normal layer has roughness, but
we assume for simplicity that the N-S
interface has a translational symmetry in the $x$-$y$ plane.

We calculate the linear response of the diamagnetic current 
to the vector potential using quasi-classical Green's function 
method proposed by Ashida {\it et al}.\cite{AAHN} and Nagato
{\it et al}.\cite{NHYN}
The vector potential is taken as ${\bf A}=(A(z),0,0)$, so that 
the magnetic field is in the $y$-direction.

We begin with the Gor'kov Green's function. According to
ref.\ \citen{AAHN}, the Gor'kov Green's function is given by a 
product of the slowly varying part and the rapidly oscillating
part with the period of the Fermi wave length. In the geometry
of present interest, the Gor'kov Green's function 
can be expanded in a form
\begin{eqnarray}
       G({\bf r},{\bf r'},\omega_n)=
       \sum_{K,K'}\sum_{\alpha=\pm}\sum_{\beta=\pm}
       G_{\alpha\beta}(K,K',z,z')
       {\rm e}^{{\rm i}K\cdot s-{\rm i}K'\cdot s'}
       {\rm e}^{{\rm i}\alpha kz-{\rm i}\beta k'z'},
\end{eqnarray}
where $\omega_n$ is the Matsubara frequency, 
$K$ is the ${\bf s}=(x,y)$ component and $k=\sqrt{2m^*E_{\rm F}-K^2}$ 
is the $z$-component of the Fermi momentum. Note that the slowly varying 
Green's functions
$G_{\alpha\beta}(K,K',z,z')$ are not diagonal in the directional
($\alpha$) space because of the
presence of the surface reflection. They are also not diagonal in $K$
space because of the surface roughness.
They
obey the first order differential
equation\cite{AAHN,NHYN}
\begin{eqnarray}
 \left({\rm i}\omega_n + {\rm i}\alpha v_K\rho_3\partial_z + 
  \frac{e}{mc}A(z)K_x-{\hat \Delta}_\alpha\right)
  G_{\alpha\beta}(K,K',z,z')=\delta_{\alpha\beta}\delta_{KK'}\delta(z-z'),
  \label{eqn:y1} 
\end{eqnarray}
where
$v_K=k/m$ is the $z$-component of the Fermi velocity, $\rho_3$ is a Pauli 
matrix in the particle-hole space and ${\hat \Delta}$ is the order parameter
matrix given by
\begin{eqnarray}
{\hat \Delta}_\alpha=
  \left(
    \begin{array}{cc}
      0&\Delta_\alpha(K,z)\\
      \Delta_\alpha^\ast(K,z)&0
    \end{array}
  \right).
\end{eqnarray}

It is convenient to introduce a one-point function
${\hat G}_{\alpha\beta}(K,K',z)$ defined by
\begin{eqnarray}
  {\hat G}_{\alpha\beta}(K,K',z)\pm {\rm i}\alpha\delta_{\alpha\beta}\delta_{KK'}=
  -2{\sqrt{v_Kv_{K'}}}\rho_3G_{\alpha\beta}(K,K',z\pm 0,z),\label{onepoint}
\end{eqnarray}
where we have taken into account the fact that 
$G_{\alpha\beta}(K,K',z,z')$ has a jump at $z=z'$ because of
the delta function in the right hand side of eq.(\ref{eqn:y1}).
The one-point function 
obeys an equation
\begin{eqnarray}
  \partial_z{\hat G}_{\alpha\beta}(K,K',z)&=&
    {\rm i}\frac{\alpha}{v_K}\hat{\varepsilon}_{K\alpha}
    \rho_3{\hat G}_{\alpha\beta}(K,K',z)-
    {\rm i}\frac{\beta}{v_{K'}}{\hat G}_{\alpha\beta}(K,K',z)
    \hat{\varepsilon}_{K'\beta}
    \rho_3,\label{eilenberger}\\
    \hat{\varepsilon}_{K\alpha}&=&
    {\rm i}\omega_n+\frac{e}{mc}A(z)K_x-\hat{\Delta}_\alpha .
    \label{eqn:y91301} 
\end{eqnarray}
When $K=K'$ and $\alpha=\beta$,  
this equation  coincides with the Eilenberger equation\cite{rf:22}.
Thus the one-point function ${\hat G}_{\alpha\beta}(K,K',z)$ 
can be interpreted as  generalized quasi-classical Green's function.

Since eq.(\ref{eilenberger}) is a first-order differential equation, 
the spatial dependence of ${\hat G}_{\alpha\beta}(K,K',z)$ can be described 
by
\begin{equation}
{\hat G}_{\alpha\beta}(K,K',z)=U_\alpha(K,z,z')
{\hat G}_{\alpha\beta}(K,K',z')U_\beta(K',z'z),\label{space}
\end{equation}
where
$U_\alpha (K,z,z')$ is the evolution operator\cite{AAHN} that
satisfies
\begin{eqnarray}
  {\rm i}v_K\partial_zU_\alpha(K,z,z')=
    -\alpha\, \hat{\varepsilon}_{K\alpha}\,
\rho_3U_\alpha(K,z,z')
\end{eqnarray}
and the initial condition 
\begin{eqnarray}
   U_\alpha(K,z,z)=1.
\end{eqnarray}

It is straightforward to calculate from eq.(\ref{eqn:y1}) the linear response 
of the Green's function $\delta G_{\alpha\beta}(K,K',z,z')$
to the vector potential $A(z)$:
\begin{eqnarray}
\hspace*{-5mm}
  \delta G_{\alpha\beta}(K,K',z,z')=
  \sum_{\gamma}\sum_{K''}\int_0^\infty{\rm d}z''
  G_{\alpha\gamma}^{(0)}(K,K'',z,z'')
  \left[-\frac{e}{mc}A(z)K_x\right]G_{\gamma\beta}^{(0)}(K'',K',z'',z'),
  \label{eqn:y2}
\end{eqnarray}
where $G_{\alpha\beta}^{(0)}$ 
is the Green's function without magnetic field. In what follows, we omit
the superscript $(0)$. The diamagnetic current is calculated from the
quasi-classical Green's function $\delta{\hat G}_{\alpha\alpha}(K,K',z)$.
Using eqs.(\ref{onepoint}) and (\ref{space}) we find\cite{higashitani}
\begin{eqnarray}
\delta{\hat G}_{\alpha\alpha}(K,K',z) & & \nonumber\\
=U_\alpha(K,z,0)\sum_\gamma&\sum_{K''}&
       \left[
       \int_0^z{\rm d}z'{\hat G}^{(+)}_{\alpha\gamma}(K,K'',0)
       \hat{\cal{L}}_{\gamma}(K'',z')
       {\hat G}^{(-)}_{\gamma\alpha}(K'',K',0)\right.\nonumber\\
       &+&\left.\int_z^\infty {\rm d}z'{\hat G}^{(-)}_{\alpha\gamma}(K,K'',0)
       \hat{\cal{L}}_{\gamma}(K'',z')
       {\hat G}^{(+)}_{\gamma\alpha}(K'',K',0)\right]U_\alpha(K',0,z),
\label{qcglinear}
\end{eqnarray}
where
\begin{eqnarray}
  \hat{G}^{(\pm)}_{\alpha\beta}(K,K',0)&=&
  \hat{G}_{\alpha\beta}(K,K',0)\pm
  {\rm i}\alpha \delta_{\alpha\beta}\delta_{KK'},\\
  \hat{\cal{L}}_{\gamma}(K,z')&=&U_\gamma(K,0,z')
  \frac{e}{mc}A(z')\frac{K_x}{v_{K}}
  \rho_3
  U_\gamma(K,z',0).
\end{eqnarray}
The quasi-classical Green's function $\delta{\hat G}_{\alpha\alpha}(K,K',z)$
is completely determined if we have a knowledge of the surface value
of ${\hat G}_{\alpha\beta}(K,K',0)$ that can be determined by solving
the boundary problem.
\section{Rough Surface Boundary Condition}

To treat the rough surface effect, we use the random $S$-matrix
theory developed by Nagato {\it et al}.\cite{NHYN,NYN} 
Let us consider the scattering of an electron at the Fermi level
by the surface. The scattering from an initial state $(K,-k)$ to 
a final state $(Q,q)$
is described by an $S$-matrix $S_{KQ}$. Since $S$ is a unitary
matrix, we can rewrite $S$ using an Hermite matrix $\eta$
\begin{equation}
S = -\ \frac{1-{\rm i}\eta}{1+{\rm i}\eta}.  \label{eqn:S}
\end{equation}
Nagato {\it et al}.\cite{NHYN,NYN} showed that one can have a formal
solution for
${\hat G}_{\alpha\beta}(K,K',0)$ that satisfies the boundary condition
described by the $S$-matrix.
The quasi-classical Green's function thus obtained 
is that for a particular configuration of the rough surface, while
the surface roughness will be distributed randomly enough over the surface.
What we are interested in is the quasi-classical Green's function
averaged over such randomness.

In what follows, we treat every element of $\eta$ as 
random variable to describe the statistical 
properties of the surface. We assume that $\eta_{KQ}$'s 
obey the Gaussian statistics with
$
\overline{\eta_{KQ}} = 0
$
and
$
\overline{\eta_{KQ}^*\eta_{K'Q'}}
= \eta^{(2)}_{KQ} \delta_{{Q}-{K}, {Q'}-{K'}}
$.

\def\bs{\overline{\sigma}}
It is instructive to discuss here on the nature of the surface
scattering in the normal state.
The average of the specular reflection amplitude is evaluated within
the self-consistent Born approximation to be\cite{NHYN,NYN}
\begin{eqnarray}
\overline{S_{{K}{K}}} &=& -\ \overline{
\left(\frac{1-{\rm i}\eta}{1+{\rm i}\eta}\right)_{{K}{K}} } 
=
 - \overline{\left( 1 - 2\eta^2 + 2\eta^4 - \cdots \right)_{KK}}
  \nonumber\\
  &=& - \left(\frac{1-\bs_{K}}{1+\bs_{K}}\right),
\label{eqn:fs}
\end{eqnarray} 
where $\bs_{K}$ obeys an integral equation
\begin{equation}
\bs_{{K}} = \sum_{Q} \eta^{(2)}_{KQ}
\frac{1}{1+\bs_{Q}}\, .
\end{equation}
In the same approximation, we can calculate the average of the scattering 
probability $\overline{|S_{KQ}|^2}$:
\begin{eqnarray}
\overline{|S_{KQ}|^2} &=& |\overline{S_{KK}}|^2\delta_{KQ} \nonumber\\
 &+& 4 
\left(\frac{1}{1+\bs_{K}}\right)^2\Gamma_{KQ}
\left(\frac{1}{1+\bs_{Q}}\right)^2,\label{scpro}
\\
\Gamma_{KQ} &=& \eta^{(2)}_{KQ} + \sum_P \eta^{(2)}_{KP}\left(\frac{1}{1+\bs_{P}}\right)^2 \Gamma_{PQ}. 
\end{eqnarray}
We adopt a simplest model and put $\eta^{(2)}_{KQ}$ constant:
\begin{equation}
\eta^{(2)}_{KQ}= \frac{2W}{\sum_{Q} 1}. \label{eqn:rough}
\end{equation}
When $W=1$, it simulates a completely diffusive surface.
The specular reflection amplitude $\overline{S_{KK}}$ vanishes
and the scattering probability $\overline{|S_{KQ}|^2}$
is independent of the outgoing momentum ${Q}$.
Nagato {\it et al}.\cite{NHYN,NYN} have shown that this model
reproduces the diffusive wall boundary condition for the Ginzburg-Landau
equation of the $p$-wave state given by 
Ambegaokar, de~Gennes and Rainer.\cite{ADR}
The case with $W=0$ corresponds to the specular surface. 
Thus, by changing $W$
from $W=0$ to $W=1$, we can discuss the surface scattering from the
specular limit to the diffusive limit. 

Let us consider the random average of 
$\delta\hat{G}_{\alpha\alpha}(K,K',z)$ of eq.(\ref{qcglinear}). 
It is convenient to write
\begin{eqnarray}
\overline{
\delta{\hat G}_{\alpha\alpha}(K,K',z)}
    =U_\alpha(K,z,0)\sum_\gamma&\sum_{K''}&
       \left[
       \int_0^z{\rm d}z'
\overline{{\hat G}^{(+)}_{\alpha\gamma}(K,K'',0)}
       \hat{\cal{L}}_{\gamma}(K'',z')
\overline{{\hat G}^{(-)}_{\gamma\alpha}(K'',K',0)}\right.\nonumber \\
    &+&
       \int_z^\infty {\rm d}z'
\overline{{\hat G}^{(-)}_{\alpha\gamma}(K,K'',0)}
       \hat{\cal{L}}_{\gamma}(K'',z')
\overline{{\hat G}^{(+)}_{\gamma\alpha}(K'',K',0)}\nonumber \\
    &+&\left.
       \int_0^\infty {\rm d}z'
\overline{X_{\alpha\gamma}(K,K'',K',z')}
       \right]
       U_\alpha(K',0,z),
\label{eqn:y0}
\end{eqnarray}
where $X_{\alpha\gamma}(K,K'',K',z')$ is defined by
\begin{eqnarray}
   X_{\alpha\gamma}(K,K'',K',z)=
   \left(
   {\hat G}_{\alpha\gamma}(K,K'',0)
   \right.
   &-&\left.\overline{{\hat G}_{\alpha\gamma}(K,K'',0)}
   ~\right)\nonumber \\
   &\times&\hat{\cal{L}}_{\gamma}(K'',0,z)
   \left(
   {\hat G}_{\gamma\alpha}(K'',K',0)
   -\overline{{\hat G}_{\gamma\alpha}(K'',K',0)}
   ~\right).
\end{eqnarray}
The average 
$\overline{{\hat G}_{\alpha\beta}(K,K',0)}$ in the random $S$-matrix
model has been given by Nagato {\it et al}.\cite{NHYN,NYN}
\begin{eqnarray}
  \overline{{\hat G}_{\alpha\beta}(K,K',0)}
  &=&{\hat G}_{\alpha\beta}(K,0)\delta_{KK'},\\
{\hat G}_{\alpha\beta}(K,0)
  &=&\left(-{\hat G}^S(K,0)^{-1}+\frac{\rm i}{2}(\alpha-\beta)\right)\nonumber \\
  &+&\left({\hat G}^S(K,0)^{-1}-{\rm i}\alpha\right)
     \frac{1}{{\hat G}^S(K,0)^{-1}-\sigma_K}
     \left({\hat G}^S(K,0)^{-1}+{\rm i}\beta\right), \label{qcg}
\end{eqnarray}
where 
${\hat G}^S(K,0)$ is the Green's function for the specular surface\cite{AAHN}
$$
{\hat G}^S(K,0)\equiv {\hat G}^S_{++}(K,0)={\hat G}^S_{--}(K,0)
$$
and 
$\sigma_K$ is the surface self energy determined by an integral equation
\begin{eqnarray}
  \sigma_K=\sum_Q\eta^{(2)}_{KQ}\frac{1}{{\hat G}^S(Q,0)^{-1}-\sigma_Q}.
  \label{sse}
\end{eqnarray}

The average of $X_{\alpha\gamma}(K,K'',K',z)$ that corresponds to
the vertex correction in the theory of impurity scattering
can be performed\cite{NYN} in the same way as in the calculation of the
scattering probability of eq.(\ref{scpro}). 
\begin{eqnarray}
\lefteqn{
 \sum_\gamma\sum_{K''}\overline{X_{\alpha\gamma}(K,K'',K',z)}
  =X_\alpha(K,z)\delta_{KK'},} \\
  && X_\alpha(K,z)=
   \left(G^S(K,0)^{-1}-{\rm i}\alpha\right)
      \left(
        Y_K\sum_Q\eta^{(2)}_{KQ}\Pi_Q(z)Y_K
      \right)
   \left(G^S(K,0)^{-1}+{\rm i}\alpha\right),
   \label{eqn:y3} 
\end{eqnarray}
where
\begin{eqnarray}
  Y_K&=&\frac{1}{{\hat G}^S(K,0)^{-1}-\sigma_K},\\
  \Pi_K(z)&=&Y_K\left(
                      {\hat M}(K,z)+\sum_Q\eta^{(2)}_{KQ}\Pi_Q(z)
                \right)
                Y_K \label{eqn:y92501},\\
  {\hat M}(K,z)&=&\sum_\gamma
                      \left(G^S(K,0)^{-1}+{\rm i}\gamma\right)
                      \hat{\cal{L}}_{\gamma}(K,0,z)
                      \left(G^S(K,0)^{-1}-{\rm i}\gamma\right).
\end{eqnarray}
Thus the averaged quasi-classical Green's function
becomes diagonal in $K$ space:
\begin{equation}
\overline{
\delta{\hat G}_{\alpha\alpha}(K,K',z)}
=\delta{\hat G}_{\alpha\alpha}(K,z)\delta_{KK'}.
\end{equation}

\section{Quasi-classical Green's Function in the Normal Layer}
In this section, we discuss the quasi-classical Green's
function in the normal layer of N-S proximity contact system.
In order to focus on the rough surface effect, we assume that
the superconducting layer is an $s$-wave semi-infinite
superconductor and
the normal layer has no pairing interaction, i.e.,
\begin{equation}
\Delta_\alpha(K,z)=
\left\{
\begin{array}{ll}0&\mbox{for}\quad 0\le z \le d, \\
\Delta(z) &\mbox{for}\quad d\le z \le \infty.
\end{array}
\right.
\end{equation}
We also adopt the simplest model of eq.(\ref{eqn:rough}) 
for surface roughness, as a result
the surface self energy $\sigma_K$ is a constant independent 
of $K$.

Under these assumptions, one finds from eq.(\ref{eqn:y92501})
that
 $\sum_Q\eta^{(2)}_{KQ}\Pi_Q(z)=0$,
because 
${\hat{\cal L}}_\gamma(K,0,z)$ is an odd function of $K$. 
It follows that $X_\alpha(K,z)=0$ and the third term in 
eq.({\ref{eqn:y0}}) can be omitted.

We further assume that the N-S interface at $z=d$ is transparent.
Then, we have for the averaged Green's function in the
normal layer
\begin{eqnarray}
  \delta{\hat G}_{\alpha\alpha}^{\rm N}(K,z)=
     U_\alpha^{\rm N}(K,z,0)&\sum_\gamma&
     \left[
     \int_0^z{\rm d}z'{\hat G}_{\alpha\gamma}^{(+)}
            {\hat{\cal L}}_\gamma^{\rm N}(K,z')
            {\hat G}_{\gamma\alpha}^{(-)}
     \right. \nonumber\\
     &+&\int_z^d{\rm d}z'{\hat G}_{\alpha\gamma}^{(-)}
            {\hat{\cal L}}_\gamma^{\rm N}(K,z')
            {\hat G}_{\gamma\alpha}^{(+)}
            \nonumber\\
     &+&\left.\int_d^\infty {\rm d}z'{\hat G}_{\alpha\gamma}^{(-)}
            {\hat{\cal L}}_\gamma^{\rm S}(K,z')
            {\hat G}_{\gamma\alpha}^{(+)}
         \right]
     U_\alpha^{\rm N}(K,0,z),
      \label{eqn:y9}
\end{eqnarray}
where
\begin{eqnarray}
  {\hat G}_{\alpha\gamma}^{(\pm)}&=&
     {\hat G}_{\alpha\gamma}(K,0)\pm {\rm i}\alpha\delta_{\alpha\gamma},\\
  {\hat{\cal L}}_\gamma^{\rm N}(K,z')&=&
     U_\gamma^{\rm N}(K,0,z')\frac{e}{mc}A(z')
     \frac{K_x}{v_K}\rho_3U_\gamma^{\rm N}(K,z',0),\\
  {\hat{\cal L}}_\gamma^{\rm S}(K,z')&=&
     U_\gamma^{\rm N}(K,0,d)U_\gamma^{\rm S}(K,d,z')\frac{e}{mc}A(z')
     \frac{K_x}{v_K}\rho_3U_\gamma^{\rm S}(K,z',d)U_\gamma^{\rm N}(K,d,0).
\end{eqnarray}
Here, $U^{\rm N(S)}$ is the evolution operator in the N~(S) layer.

The explicit form of the Green's function ${\hat G}_{++}$ 
is given by\cite{AAHN,NHYN}
\begin{eqnarray}
  {\hat G}_{++}(K,0)={\rm i}\left(
                     \frac{\tilde A}{\frac{1}{2}{\rm Tr}{\tilde A}}-1
                     \right),
\end{eqnarray}
where
\begin{eqnarray}
  {\tilde A}=
      U_+^{\rm N}(K,0,d)U_+^{\rm S}(K,d,\infty)U_-^{\rm S}(K,\infty,d)U_-^{\rm N}(K,d,0)
      \frac{1-{\rm i}\sigma}{1+{\rm i}\sigma}
      \label{eqn:y4}
\end{eqnarray}
and {\rm Tr} denotes the trace in particle-hole space. We note that
the surface self energy $\sigma_K=\sigma$ is $K$-independent in the
present model of eq.(\ref{eqn:rough}).
The $U^{\rm S}_{\pm}(z,z')$ carries all the information on the
superconducting layer. It depends in general 
on the spatial profile of the order parameter $\Delta(z)$.
But we assume for simplicity that $\Delta(z)$ is a constant
and is equal to the bulk BCS value $\Delta(T)$.
By this simplification, we can have an analytic solution
for $U^{\rm S}_{\pm}(z,z')$ as well as for $U^{\rm N}_{\pm}(z,z')$:
\begin{eqnarray}
    U^{\rm N}_\alpha(z,z')&=&\exp\left[-\frac{\alpha}
    {v_{{\rm F}_z}}{\omega_n}\rho_3(z-z')\right],\\
    U^{\rm S}_\alpha(z,z')&=&\exp\left[-\frac{\alpha}{v_{{\rm F}_z}}
    \left(\omega_n\rho_3+\Delta\rho_2\right)(z-z')\right].
\end{eqnarray}
Substituting them into eq.(\ref{eqn:y4}), we find
\begin{eqnarray}
  {\tilde A}=\frac{1}{2\Omega}
             \left(
               \begin{array}{cc}
                (\Omega+\omega_n){\rm e}^{\kappa d} & -{\rm i}\Delta\\
                {\rm i}\Delta & (\Omega-\omega_n){\rm e}^{-\kappa d}
               \end{array}
             \right)
             \frac{1-{\rm i}\sigma}{1+{\rm i}\sigma},
\end{eqnarray}
where $\Omega=\sqrt{\omega_n^2+\Delta^2}$ and $\kappa=2\omega_n/v_{{\rm F}_z}$.
The surface self energy is determined from eq.(\ref{sse}).
This has to be done numerically for each Matsubara frequency
$\omega_n$.

The other Green's functions 
$\hat{G}_{+-},\hat{G}_{-+}$ and $\hat{G}_{--}$ 
are related to $\hat{G}_{++}$ via eq.(\ref{qcg}).

\section{Diamagnetic Current}

In this section, we calculate the diamagnetic current using the 
quasi-classical Green's function derived in the previous section. 
The diamagnetic 
current $j(z)$ which flows in the $x$-direction in the present 
geometry is given by
\begin{eqnarray}
  j(z)&=&(-e)T\sum_{\omega_n}\sum_Kv_{{\rm F}_x}{\rm Tr}~\rho_3
       \sum_\alpha\delta{\hat G}_{\alpha\alpha}(K,z)\nonumber\\
  &=&-e\left(\pi N(0)/2\right)
       T\sum_{\omega_n}\int_0^{\pi/2}{\rm d}\theta\sin\theta
       \int_0^{2\pi}\frac{{\rm d}\varphi}{2\pi}
       v_{{\rm F}_x}{\rm Tr}~\rho_3
       \sum_\alpha\delta{\hat G}_{\alpha\alpha}(K,z),
\end{eqnarray}
where $N(0)$ is the density of states at the Fermi level, 
$\theta$ is the polar angle of the Fermi momentum and
$v_{{\rm F}_x}=v_{\rm F}\sin\theta\cos\varphi$ is the
$x$-component of the Fermi velocity.

As was shown by Zaikin\cite{zaikin} and Higashitani and 
Nagai\cite{higashitani}, the diamagnetic current $j(z)$
is constant throughout the normal layer and the 
current-field relation is completely nonlocal.
This is characteristic to the clean system and happens because
the normal state evolution operator $U_\alpha^{\rm N}(K,z,z')$
commutes with $\rho_3$, therefore 
Tr$\rho_3\sum_\alpha \delta G_{\alpha\alpha}^{\rm N}(K,z)$ 
is independent of $z$.

We show below  that rough surface
effect is large even if the normal layer width $d$ is
larger than the coherence length $\xi$ and the penetration depth
$\lambda$. In such a case, we can neglect the 
third term in eq.(\ref{eqn:y9}), 
because the vector potential $A(z)$  decays exponentially.
Now we have an expression for the diamagnetic current
in the normal layer:
\begin{eqnarray}
  j(z)=-\frac{e^2}{mc}n_{\rm N}^*{\bar A},
                   \label{eqn:y5}
\end{eqnarray}
where ${\bar A}$ is the vector potential averaged over the normal 
layer and
\begin{eqnarray}
  n_{\rm N}^*=n_{\rm N}d\pi T\sum_{\omega_n}
  \langle\!\langle{\cal K}/v_{{\rm F}_z}\rangle\!\rangle
   \label{eqn:y1017}
\end{eqnarray}
is the effective superfluid density, i.e., the density of 
proximity-induced superconducting electrons in the normal layer.
Here,
\begin{eqnarray}
  {\cal K}=\frac{1}{2}{\rm Tr}~\rho_3\sum_{\alpha\gamma}
           {\hat G}_{\alpha\gamma}^{(+)}\rho_3
           {\hat G}_{\gamma\alpha}^{(-)},
\end{eqnarray}
$n_{\rm N}=p_{\rm F}^3/3\pi^2$ is the electron number density in the 
normal layer and the double bracket means the angle average
over the polar angle
\begin{eqnarray}
  \langle\!\langle~\cdots~\rangle\!\rangle=
  \frac{3}{2}\int_0^{\pi/2}{\rm d}\theta\sin^3\theta~(~\cdots~).
\end{eqnarray}
Combining eq.(\ref{eqn:y5})  with the Maxwell equation
$j_x=-\frac{c}{4\pi}\partial^2_zA(z)$, we find the magnetic field 
$B(z)$ in the normal layer when the applied external magnetic field
is $H$:
\begin{eqnarray}
  B(z)=H\left(1-z/\lambda\right),
       \label{eqn:y8}
\end{eqnarray}
where
\begin{eqnarray}
  \lambda=d\left(\frac{2\lambda^2_{\rm NL}n_{\rm N}}{d^2n_{\rm N}^*}+\frac{2}{3}\right)
       \label{eqn:y6}
\end{eqnarray}
and $\lambda_{\rm NL}$ is the London penetration depth of the normal 
layer defined by
\begin{eqnarray}
  \lambda_{\rm NL}=\sqrt{\frac{mc^2}{4\pi e^2n_{\rm N}}}.
\end{eqnarray}

In Fig.~\ref{fig:nN}, we show the surface roughness dependence of
$n_{\rm N}^*/n_{\rm N}$ for $d/\xi$=20 and 50 ($\xi=v_{\rm F}/\pi T_{\rm c}$ 
the coherence length). 
The solid curve represents the temperature dependence of $n_{\rm N}^*/n_{\rm N}$
in the diffusive limit and the dashed one represents the specular 
limit.\cite{higashitani}
In both the limits $n_{\rm N}^*$ shows similar temperature dependence but 
the magnitude is considerably different. As was shown 
in ref.\ \citen{higashitani}, $n_{\rm N}^*$ is exponentially small 
at high temperatures and begins to grow below temperature 
$\sim (\xi/d)T_{\rm c}$  at which the order parameter extends 
up to the normal layer end. 
The magnitude of $n_{\rm N}^*$ is strongly reduced, in particular, 
at low temperatures by the diffusive scattering effect. 
The largest reduction is obtained at $T=0$. We see that 
$n_{\rm N}^*/n_{\rm N}$ at $T=0$ reaches to unity in the specular case but 
is reduced to about 0.65 in the diffusive limit. 
We can also see that the suppression 
rate of $n_{\rm N}^*/n_{\rm N}$ does not depend upon the normal layer width $d$. 
This means that, at low temperatures, the rough surface effect 
has to be taken into account even in such thick normal layers 
that satisfy the conditions $d > \xi$ or $d > \lambda_{\rm NL}$.

\section{Screening Length}
In this section, we discuss the screening length $\rho$  defined 
by
\begin{eqnarray}
\rho=\frac{1}{H}\int_0^d {\rm d}z \left(H-B(z)\right).
\label{eqn:y7}
\end{eqnarray}
This length is equivalent to total diamagnetic susceptibility 
of the normal layer. 
The explicit expression for the screening length $\rho$ in clean N-S systems 
was first given in ref.\ \citen{higashitani}. 
The same result holds for the present N-S system and is given by
\begin{eqnarray}
\rho/d=\frac{3}{4}\frac{1}{1+3(\lambda_{\rm NL}/d)^2(n_{\rm N}/n_{\rm N}^*)}.
\label{scrlength}
\end{eqnarray}

Figure \ref{fig:sl} shows the temperature dependence of the 
screening fraction $\rho/d$ when $d/\xi=20$ and $d/\xi=50$. 
The solid curves represent the diffusive limit and the dashed curves 
represent the specular limit. 
The temperature dependence of $\rho/d$ comes only from $n_{\rm N}^*$. 
When $n_{\rm N}^*/n_{\rm N} \ll 1$ at high temperatures, 
the screening fraction can be approximated by 
$\rho/d \simeq \frac{1}{4}(n_{\rm N}^*/n_{\rm N})/(\lambda_{\rm NL}/d)^2$ and 
is proportional to the effective superfluid density $n_{\rm N}^*$. 
Accordingly, $\rho/d$ increases exponentially with decreasing temperature. 
At low temperatures, since $n_{\rm N}^*/n_{\rm N}$ becomes of order unity 
and we are considering thick normal layers such that satisfy 
$\lambda_{\rm NL}/d \ll 1$, we may neglect the term 
$3(\lambda_{\rm NL}/d)^2(n_{\rm N}/n_{\rm N}^*)$ 
in the denominator of eq.(\ref{scrlength}), so that 
$\rho/d$ exhibits saturation to $3/4$. 
These high and low temperature behaviors of $\rho/d$ are common to 
the specular and the diffusive limits. 
The difference between the two results is therefore visible only in the 
intermediate temperature range, though the superfluid density 
$n_{\rm N}^*$ has significant difference at all temperatures. 

Belzig {\it et al}.\cite{belzigb,belzigc} discussed the impurity 
scattering effect on $\rho/d$ in a same system with our's
but with specular surface. Hara {\it et al}.\cite{hara} considered
the effect by finite reflection at the interface in addition
to the impurity effect. The finite reflection has a similar effect 
to the rough surface effect, namely, that dose not 
alter much the qualitative temperature dependence of $\rho$ but reduces 
considerably its magnitude. 
The impurity effect is more involved. In dirty systems, 
the screening fraction $\rho/d$ increases rather slowly 
with decreasing temperature ($\rho \propto T^{-1/2}$ in the dirty limit) 
and reaches to a larger value than the clean limit one $3/4$ at $T=0$. 
The detailed structure of the temperature dependence of $\rho$ 
is sensitive to the impurity concentration.\cite{belzigb} 
The difference in $\rho$ between 
clean and dirty systems originates from the fact that the impurity 
scattering makes the magnetic response be local.\cite{zaikin,higashitani}

When compared with the results of Belzig {\it et al}.\ \cite{belzigb,belzigc}
and Hara {\it et al}.\ \cite{hara}, 
the present analysis suggests that the rough surface scattering effect is
quantitatively as large as those by impurity scattering
and by finite reflection at the interface.
This means that when a system is comparatively clean we have to 
take into account the rough surface effect simultaneously 
with those by the impurity scattering and by the finite reflection
at the interface.

\section{Summary and Discussion}
We have studied the rough surface effect on the diamagnetic current 
in the N-S proximity-contact system using the quasi-classical theory. 
The linear response of the quasi-classical Green's function
to the vector potential has been calculated 
correctly taking into account the rough surface effect.

We studied the temperature dependence of the diamagnetic current 
in the clean normal layer and 
found that the diamagnetic current is significantly suppressed 
by the diffusive surface scattering even in a thick normal 
layer of which the width is larger than
the coherence length $\xi=v_{\rm F}/\pi T_{\rm c}$. 
We have shown that in comparatively clean systems the
rough surface scattering effect is as significant as
those by the impurity scattering and by the finite reflection
at the interface. Those effects should be considered 
simultaneously for the quantitative comparison with
experiments.

Before concluding this report it is worth mentioning
the paramagnetic re-entrant behavior of a N-S proximity
contact system reported by Mota and co-workers.\cite{mota,visani}
In the present calculation, no such behavior was found.
Bruder and Imry\cite{bi} suggested that the 
glancing current 
along the surface of normal metal cylinder 
contributes to the paramagnetic
susceptibility.
But the theory was criticized for the predicted paramagnetism 
being too small to explain the experiment.\cite{fau}
Fauch\` ere {\it et al.}~\cite{fbb} pointed out 
that when the pairing 
interaction in the normal metal is repulsive, a zero energy
bound state is formed at the interface and that zero-energy
bound state contributes to the paramagnetic current.\cite{hp}
The bound state energy, however, is not always equal to
zero  but depends on the reflection coefficient $R$ at the
interface.\cite{nn}
More recently, Maki and Haas\cite{mh} proposed that 
the normal metals 
(noble metals) may become $p$-wave superconductor and generate a 
counter-current in the normal metal.
The problem of the paramagnetic re-entrant behavior 
is still to be studied theoretically as well as experimentally.

\section*{Acknowledgements}
We would like to thank D.~Rainer, J.~Hara, M.~Ashida, Y.~Nagato and 
M.~Yamamoto for helpful discussions.
This work is supported in part by a Grant-in-Aid for Scientific Research
(No.11640354) and a Grant-in-Aid for COE
Research (No.13CE2002) of the Ministry of Education, Culture,
Sports, Science and Technology of Japan.

\newpage

\begin{figure}
\figureheight{0cm}
\caption{A geometry of N-S double layer system with a rough surface.
The solid and the dashed curves represent the order parameter
on the S side and on the N side, respectively.
}
\label{fig:NS}
\end{figure}

\vspace{-3cm}

\begin{figure}
\figureheight{0cm}
\caption{The effective superfluid density $n^*_{\rm N}$ in the 
normal layer as a function of temperature $T/T_{\rm c}$ for layer 
widths $d/\xi=20$ and $d/\xi=50$. 
Solid curves represent the diffusive limit. 
Dashed curves represent the specular limit.
}
\label{fig:nN}
\end{figure}

\vspace{-3cm}

\begin{figure}
\figureheight{0cm}
\caption{Temperature dependence of screening length $\rho$ for layer 
widths $d/\xi=20$ and $d/\xi=50$. 
Solid curves represent the diffusive limit. 
Dashed curves represent the specular limit.
}
\label{fig:sl}
\end{figure}

\newpage
\pagestyle{empty}
\hspace*{-1.5cm}
\epsfile{file=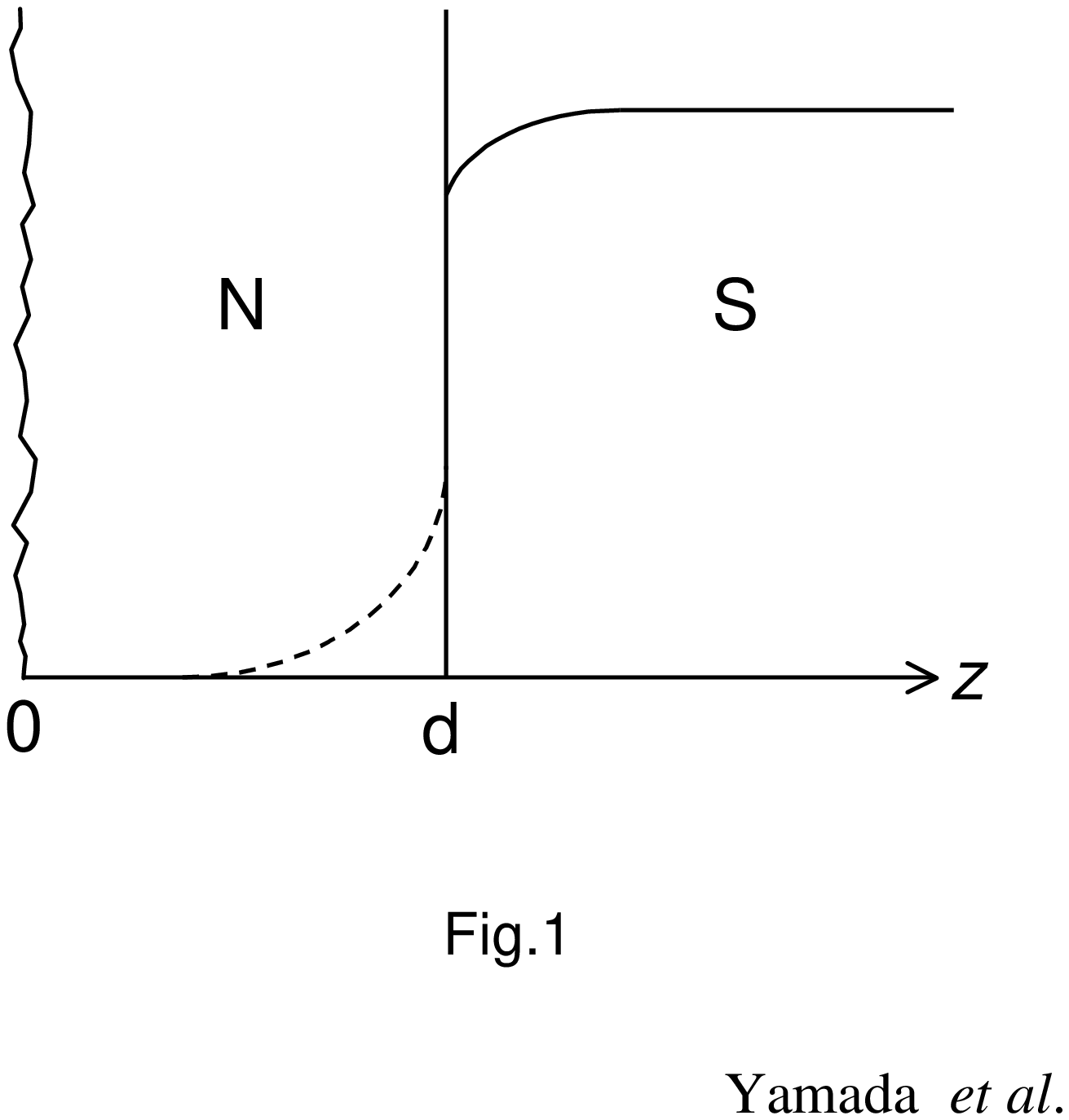}

\newpage
\pagestyle{empty}
\hspace*{-1.5cm}
\epsfile{file=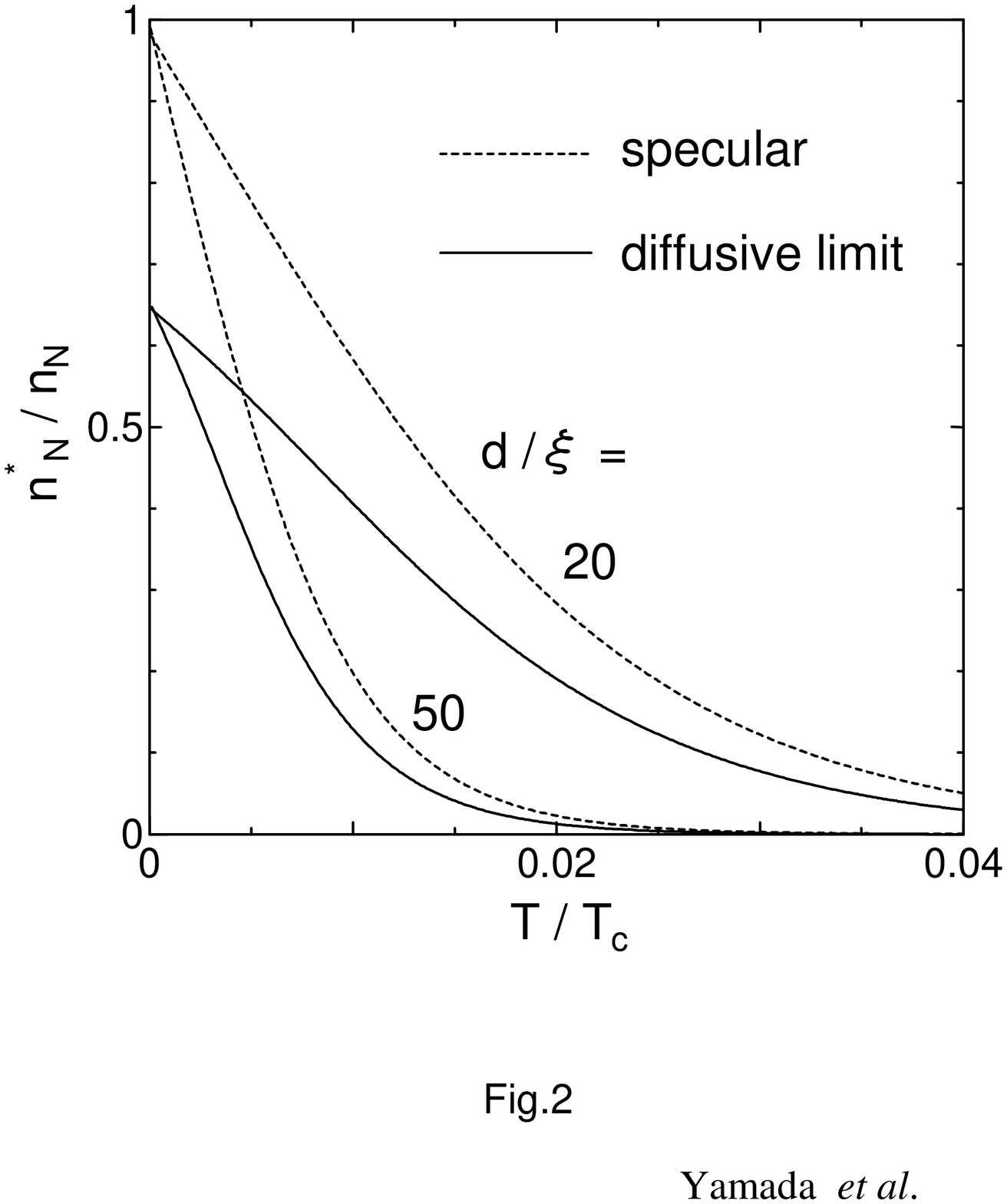}

\newpage
\pagestyle{empty}
\hspace*{-1.5cm}
\epsfile{file=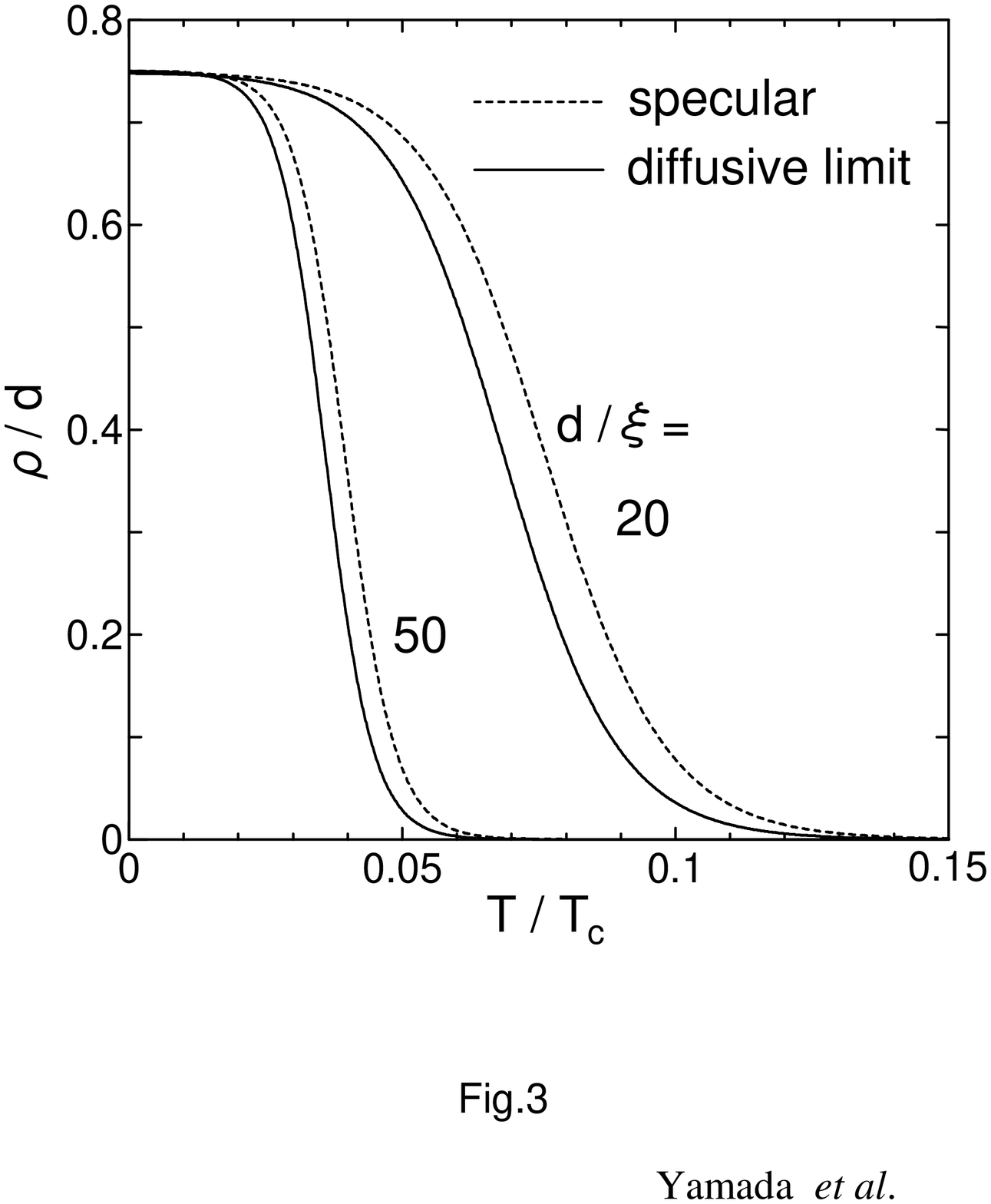}

\end{document}